\documentclass[aps,reprint,twocolumn,groupedaddress]{revtex4-2}
\usepackage{graphicx}
\usepackage{dcolumn}
\def\Vec#1{\mbox{\boldmath $#1$}}

\begin{document}


\title{A Rotation-Compensated Smartphone Accelerometer Application\\for Undergraduate Mechanics Experiments}


\author{Keita Nishioka}
\email[Corresponding author: ]{knisi@neptune.kanazawa-it.ac.jp}
\affiliation{Math and Science Education Research Center, Kanazawa Institute of Technology, Japan}

\author{Yasuhiro Tanaka}
\affiliation{Math and Science Education Research Center, Kanazawa Institute of Technology, Japan}



\begin{abstract}
Smartphones equipped with sensors such as accelerometers, gyroscopes, and magnetometers offer valuable opportunities for physics education, allowing students to measure motion using their own devices. However, commonly used applications provide acceleration only in the device-fixed coordinate system, which makes it difficult to analyze two- or three-dimensional motion when the device rotates. To address this limitation, we developed a web-based accelerometer application that can provide acceleration in a stationary global coordinate system. This is achieved by simultaneously recording acceleration in the device-fixed coordinate system and  Euler angles, and converting them to rotation-compensated acceleration in real time. We also built a companion web application for numerical integration, noise reduction, and visualization of the measured data. Both applications are installation-free and can be accessed directly through a smartphone browser.

We demonstrate the capabilities of the newly developed system through several representative types of motion, including sliding motion, projectile motion, and circular motion, by showing that rotation-compensated acceleration enables accurate reconstruction of velocity, displacement, and trajectories even when the smartphone changes its orientation. The applications were implemented in undergraduate mechanics classes, where students used them in group-based experiments. Classroom observations suggested that the use of these tools facilitated a deeper understanding of the relationships among acceleration, velocity, and position. These results suggest that rotation‑compensated smartphone measurements provide a practical and effective tool for physics education.
\end{abstract}

\keywords{Smartphone sensors, Acceleration measurement, Rotation compensation, Physics education, Mechanics experiment}

\maketitle
\section{Introduction}
The increasing availability of sensors in modern smartphones has created new opportunities for physics education. Equipped with accelerometers, gyroscopes, magnetometers, and microphones, smartphones allow students to perform experiments using devices they own, reducing cost barriers and often motivating students to engage with familiar technology in novel ways.

Early demonstrations of this potential include experimental studies of the physical pendulum using built-in acceleration and gyroscope sensors, which enabled direct measurement of radial and tangential accelerations, angular velocity, and even phase-space trajectories \cite{Monteiro2014}. Since then, numerous studies have explored smartphone-based experiments in physics teaching. Applications such as phyphox and Physics Toolbox Sensor Suite provide versatile platforms for data acquisition and real-time visualization across Android and iOS devices \cite{Staacks2018,Vieyra2015}. These tools have supported experiments ranging from time-of-flight measurements of sound \cite{Staacks2019} and demonstrations of Bernoulli's principle \cite{Macchia2017} to gamified learning activities \cite{Vieyra2020}, large-scale collaborative experiments \cite{Staacks2022}, and magnetic oscillation measurements in LC circuits \cite{Westermann2022}. We have also reported acceleration-measurement activities conducted in introductory mechanics courses for second-year university students using conventional smartphone measurement applications \cite{Nishioka2020}. Together, these studies highlight smartphones as cost-effective, sensor-rich instruments for physics education.

Despite these advantages, most commonly used applications provide acceleration only in the device-fixed coordinate system. Consequently, analysis is typically limited to one-dimensional motion with fixed device orientation. When the smartphone rotates—an almost inevitable situation in two- and three-dimensional motions such as projectile or circular motion—the device axes rotate relative to the stationary global frame. Direct integration of device-frame acceleration in such cases produces physically inconsistent velocity and position data, limiting meaningful motion analysis.

To overcome this limitation, we prototyped a web-based acceleration-measurement application that records device-frame acceleration together with Euler angles and transforms the data into a stationary global coordinate system in real time \cite{Nishioka2023}. This rotation compensation enables consistent reconstruction of acceleration regardless of device orientation. Because the system operates entirely within a web browser, it requires no installation and is accessible on any smartphone with an internet connection. We also developed a companion browser-based analysis tool that performs numerical integration, filtering, and visualization, allowing students to obtain velocity, position, and trajectories without configuring spreadsheet software. To our knowledge, no existing smartphone application combines real-time rotation compensation in a global frame with integrated browser-based numerical analysis.

In this paper, we describe the design and implementation of these applications, present representative measurement examples, and report their use in undergraduate mechanics classes. By enabling rotation-compensated acceleration measurements and streamlined analysis workflows, the system expands the scope of smartphone-based experiments and supports students' understanding of motion in one, two, and three dimensions.

\section{Development of Rotation-Compensated Acceleration Measurement Application}

\subsection{Sensor APIs and Orientation Acquisition}
\begin{figure*}
  \begin{center}
    \includegraphics[width=0.7\linewidth]{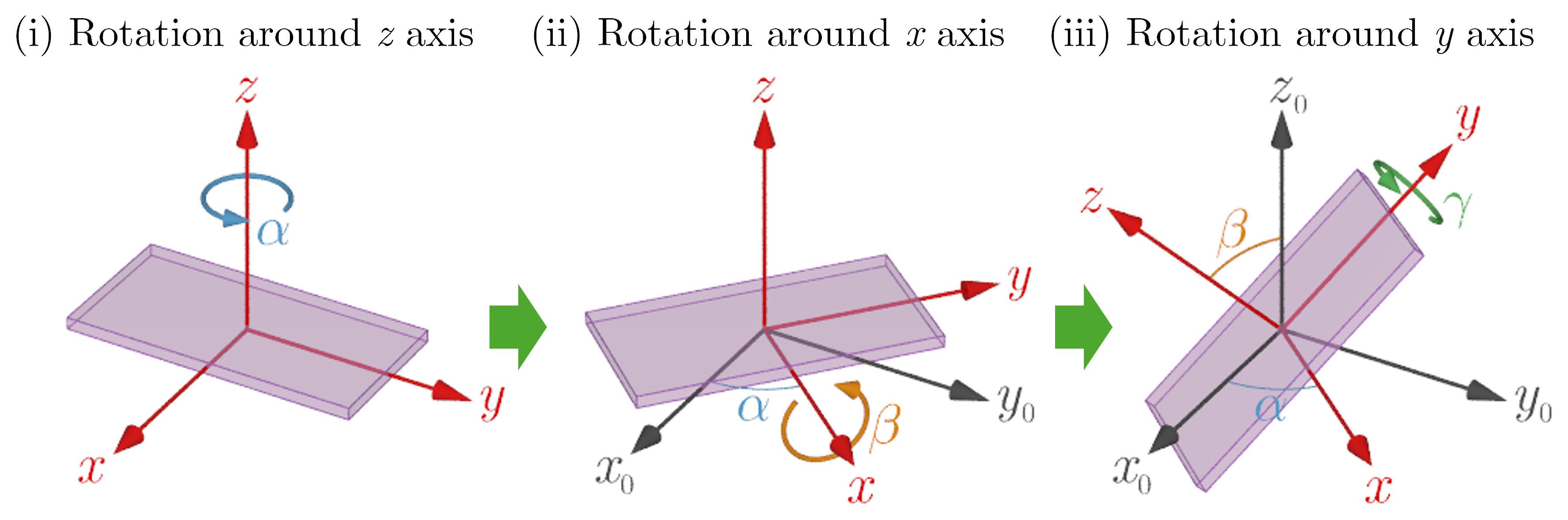}
    \caption{Order of rotation of $z$-$x$-$y$ Euler angles. (i), (ii) and (iii) are the schematic representations of rotation of a mobile device around $z$, $x$ and $y$ axes, respectively. The light purple plate represents the device such as a smartphone.}
    \label{fig:z-x-y_euler_angle}
  \end{center}
\end{figure*}

To access motion and orientation data, web browsers utilize two primary interfaces (APIs), the DeviceMotionEvent and the DeviceOrientationEvent. The former obtains the device's acceleration and rotational velocity, while the latter obtains the device's orientation in terms of Euler angles. As shown in Fig. \ref{fig:z-x-y_euler_angle}, the Euler angles $\alpha$, $\beta$, and $\gamma$ obtained from the DeviceOrientationEvent API represent sequential rotations about the device-fixed $z$, $x$, and $y$ axes, respectively, that is, they correspond to $z$–$x$–$y$ Euler angles. These angles can be converted into rotation matrices suitable for coordinate transformation. Note that this convention differs from the roll–pitch–yaw angles commonly used for vehicles and robots, which follow a $z$–$y$–$x$ Euler angle sequence.

\subsection{Coordinate Transformation}\label{sec:coordinate_transformation}
\begin{figure}
  \begin{center}
    \includegraphics[width=0.85\linewidth]{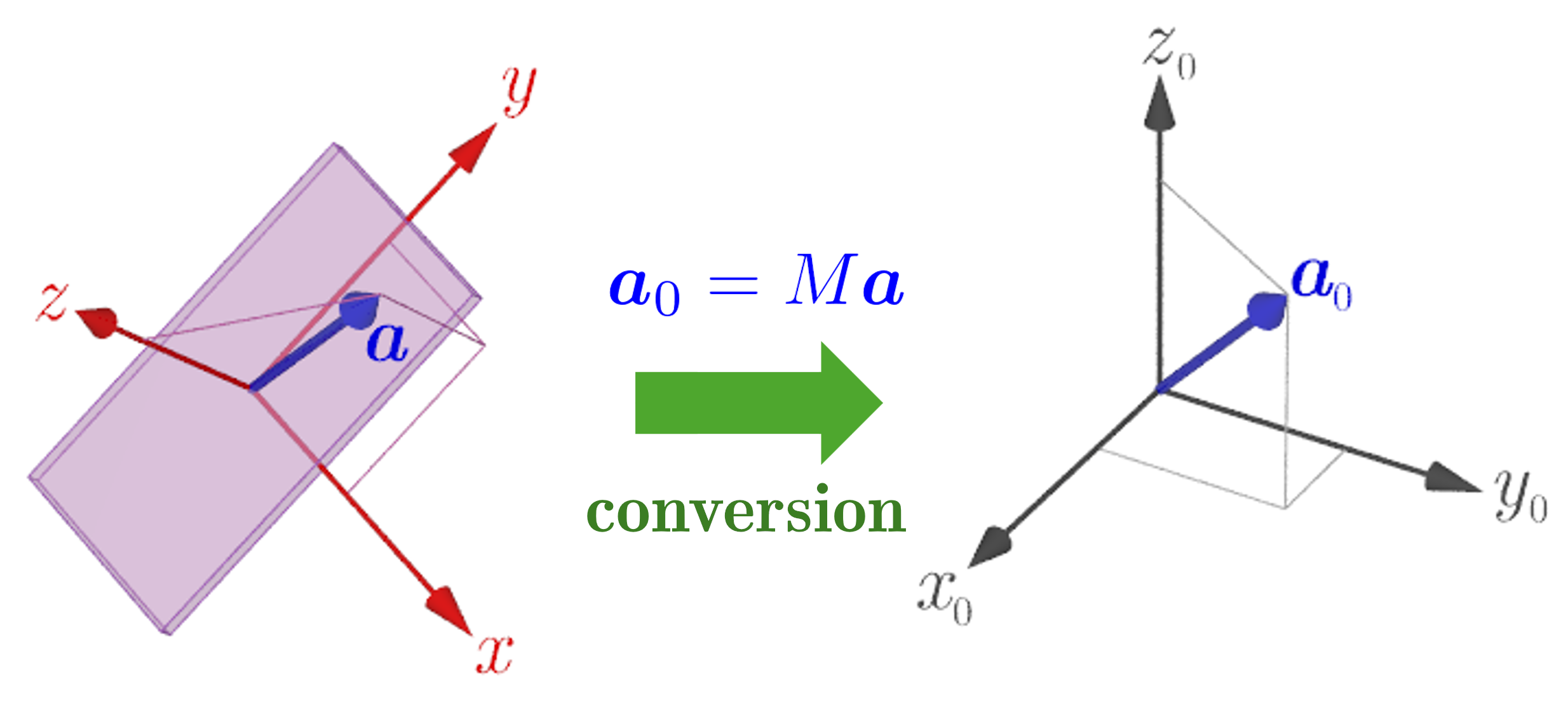}
    \caption{Conversion of acceleration from a device-fixed coordinate system to a stationary one.}
    \label{fig:acceleration_conversion}
  \end{center}
\end{figure}

Acceleration measured in the device-fixed coordinate system is transformed into that in a stationary global coordinate system. In this global system, the $z$ axis corresponds to the upward vertical direction, and the $xy$ plane represents the horizontal plane, whose orientation is initialized at the start of the measurement. Measuring the Euler angles alongside acceleration data allows for the determination of the device's orientation relative to the initial stationary frame. The rotation matrices $R_z(\alpha)$, $R_x(\beta)$, $R_y(\gamma)$ for the Euler angles $\alpha$, $\beta$, $\gamma$ are given by
\begin{eqnarray}
  R_z(\alpha)&=&\left(
    \begin{array}{ccc}
      \cos\alpha & -\sin\alpha & 0 \\
      \sin\alpha & \cos\alpha & 0 \\
      0 & 0 & 1
    \end{array}
  \right),\\
  R_x(\beta)&=&\left(
    \begin{array}{ccc}
      1 & 0 & 0 \\
      0 & \cos\beta & -\sin\beta \\
      0 & \sin\beta & \cos\beta
    \end{array}
  \right),\\
  R_y(\gamma)&=&\left(
    \begin{array}{ccc}
      \cos\gamma & 0 & \sin\gamma \\
      0 & 1 & 0 \\
      -\sin\gamma & 0 & \cos\gamma
    \end{array}
  \right).
\end{eqnarray}
Consider converting $\Vec{a}_0=(a_{0x},\ a_{0y},\ a_{0z})$ expressed in the stationary coordinate system into $\Vec{a}=(a_x,\ a_y,\ a_z)$ in a coordinate system rotated by the $z$-$x$-$y$ Euler angles. In this case, $\Vec{a}$ is obtained by rotating $\Vec{a}_0$ sequentially backward around the $z$ axis by $\alpha$, the $x$ axis by $\beta$, and the $y$ axis by $\gamma$. This relationship is expressed as
\begin{eqnarray}
\Vec{a}&=&R_y^{-1}(\gamma)R_x^{-1}(\beta)R_z^{-1}(\alpha)\Vec{a}_0\nonumber\\
&=&\left\{R_z(\alpha)R_x(\beta)R_y(\gamma)\right\}^{-1}\Vec{a}_0\nonumber\\
&=&M^{-1}\Vec{a}_0,\label{eq:acceleration_in_device-fixed_system}
\end{eqnarray}
where $M=R_z(\alpha)R_x(\beta)R_y(\gamma)$. In practice, since the DeviceMotionEvent API provides the device-frame acceleration $\Vec{a}$, applying the inverse transformation of eq.(\ref{eq:acceleration_in_device-fixed_system}), $\Vec{a}_0=M\Vec{a}$, yields the rotation-compensated acceleration in the stationary global frame, even when the smartphone's orientation changes during motion (Fig. \ref{fig:acceleration_conversion}).

\subsection{Implementation of the Web Application}
\begin{figure}
  \begin{center}
    \includegraphics[width=0.8\linewidth]{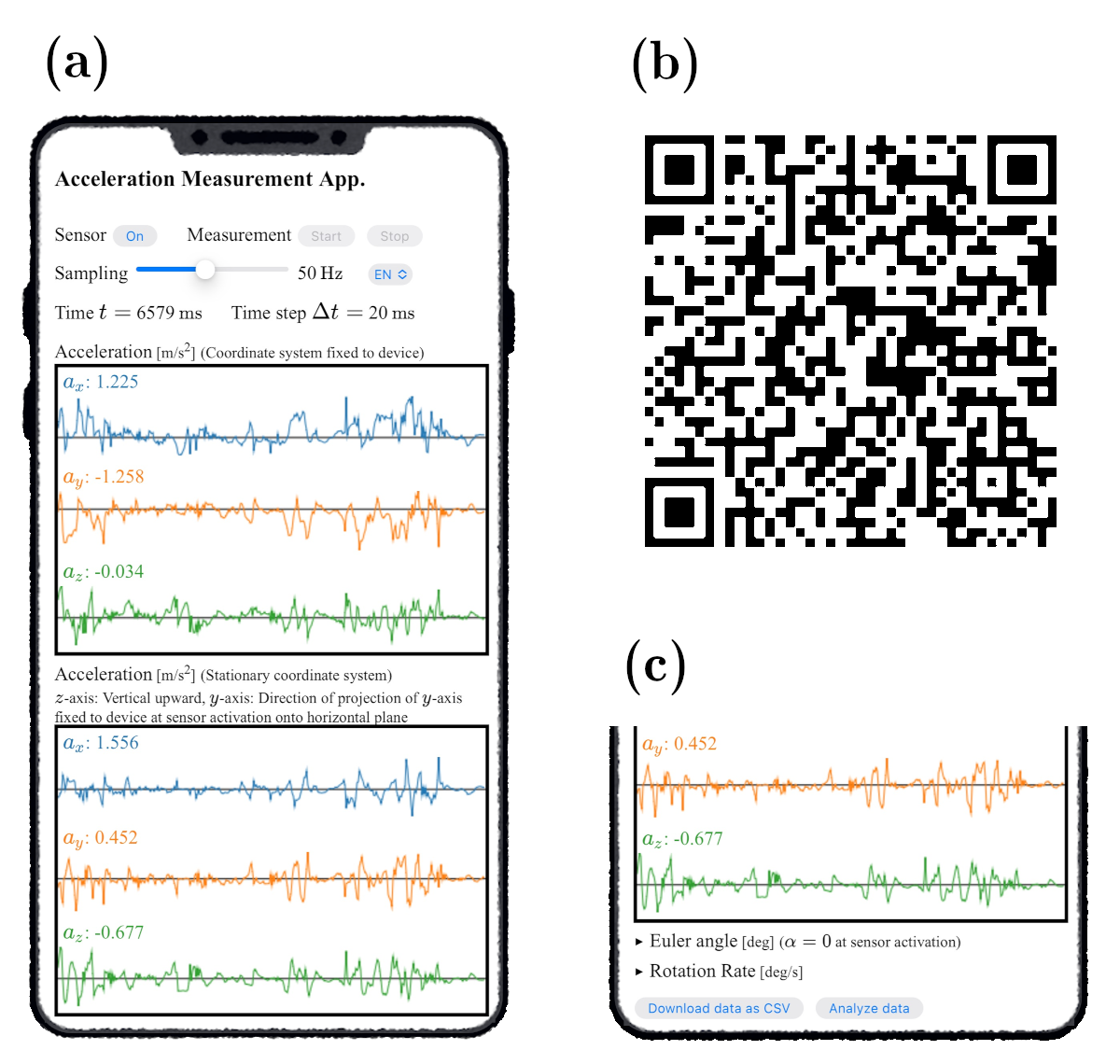}
    \caption{(a) Screen of acceleration measurement app. (b) Two-dimensional code of the measurement app. (c) Bottom area of this application.}
    \label{fig:acceleration_measurement_app}
  \end{center}
\end{figure}

Based on the rotation compensation scheme described in Section \ref{sec:coordinate_transformation}, we implemented a web-based acceleration measurement application that performs real-time coordinate transformation using simultaneously acquired Euler angles. Figure \ref{fig:acceleration_measurement_app}(a) shows the screen of this application, which is publicly available online and can be accessed directly via a smartphone browser, either by scanning the two-dimensional code shown in Fig. \ref{fig:acceleration_measurement_app}(b) or by visiting the project website (https://natieck.github.io/phys\_apps/) and selecting the ``Acceleration Measurement App" link. As the system is entirely web-based, it requires no installation or platform-specific configuration, ensuring immediate accessibility across various smartphones.

The application acquires acceleration and orientation data in parallel and applies the rotation compensation in real time. This allows users to monitor both the raw device-frame acceleration and the transformed global-frame acceleration during measurement. Such direct visualization clarifies how device rotation affects the measured acceleration and how the compensation restores physically meaningful components in a stationary global frame.

Measured data can be downloaded as a CSV file containing acceleration in both coordinate systems, Euler angles, and angular velocities, and can be analyzed using standard spreadsheet software. As shown in Fig. \ref{fig:acceleration_measurement_app}(c), the measurement interface includes two buttons at the bottom of the screen: a ``Download data as CSV" button and an ``Analyze data" button. The former allows users to save recorded measurement data as a CSV file, while the latter transfers the data directly to a companion web-based analysis application, where numerical integration and visualization of velocity, position, and trajectories can be performed immediately within the browser environment.

Implementing the system as a web application offers several pedagogical advantages. First, it eliminates installation barriers and operating-system dependencies, which is particularly important in classroom settings where students use heterogeneous devices. Second, updates and improvements are instantly available to all users without requiring manual upgrades. Finally, the seamless integration of measurement and analysis within a browser supports efficient experimentation and interpretation within limited class time.

This application has been tested on both iOS and Android devices, but slight variations in measurement accuracy were observed depending on the device and browser.

\section{Development of Web-Based Analysis Application}
In our previous acceleration measurement experiments, students analyzed recorded data by manually entering formulas into spreadsheet software, such as Microsoft Excel, to perform numerical integration. However, we observed that some students unfamiliar with spreadsheet-based calculations often were unable to obtain correct velocity and position data due to unnoticed errors in formula entry. Consequently, these students frequently lost track of the analytical objectives and struggled to grasp the underlying physical principles of the calculations \cite{Nishioka2020,Nishioka2023}. Such difficulties indicate that excessive time spent on non-essential technical procedures can hinder conceptual learning in physics experiments.

\begin{figure}
  \begin{center}
    \includegraphics[width=1.0\linewidth]{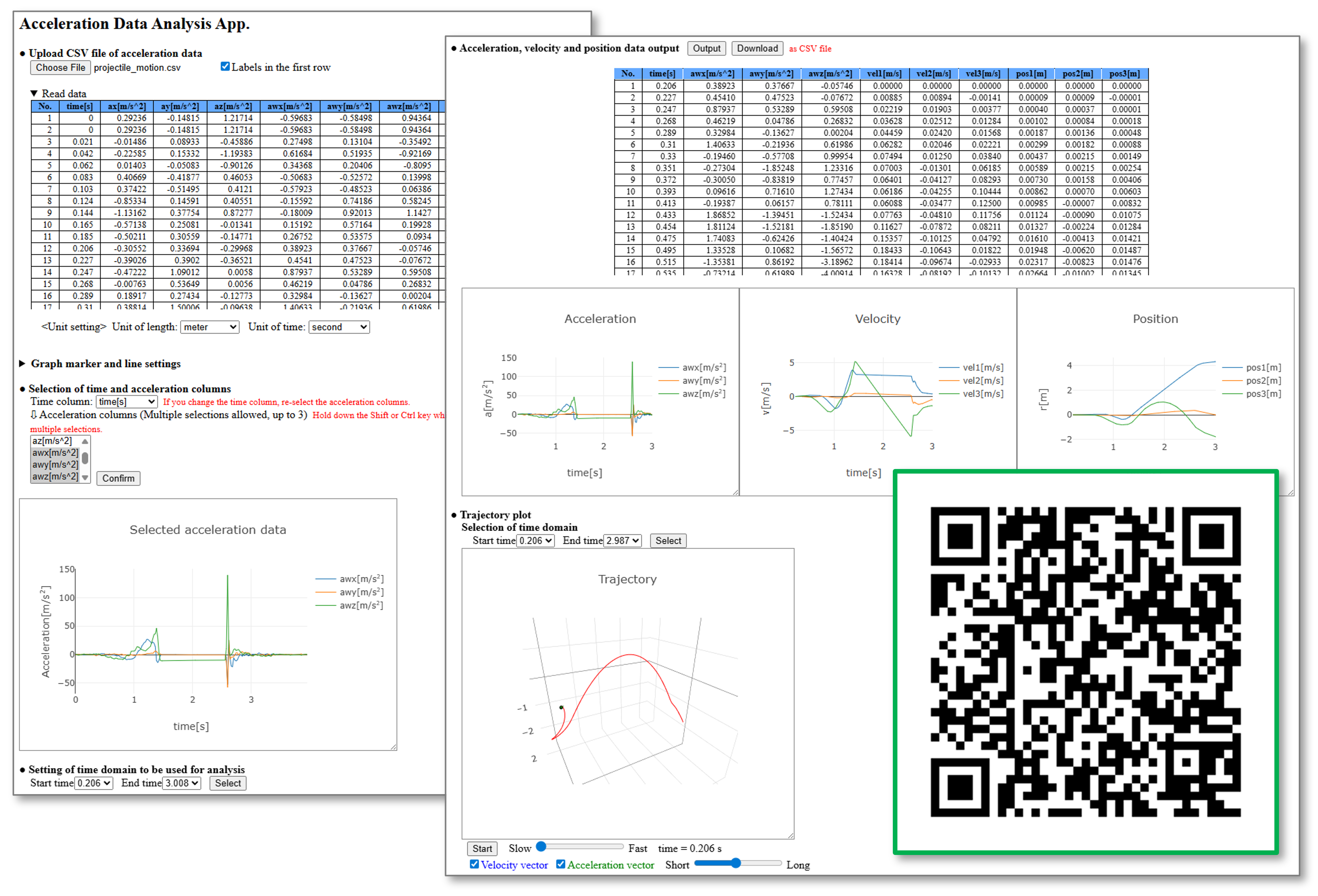}
    \caption{Screen of acceleration data analysis app. The inset at the right bottom is the two-dimensional code of the manual page for the analysis app.}
    \label{fig:acceleration_data_analysis_app}
  \end{center}
\end{figure}

To address this issue, we developed a web-based analysis application that automatically processes measured acceleration data and computes velocity and position through numerical integration (Fig. \ref{fig:acceleration_data_analysis_app}). By simply loading the acceleration data file obtained from the measurement application, students can immediately analyze the data without manually entering formulas. This design allows them to focus on interpreting the physical outcomes rather than on spreadsheet operations.

The analysis application includes several integrated functions that support effective data interpretation. It features noise-reduction capabilities to suppress sensor-related fluctuations, thereby improving the reliability of the numerical integration. The application also provides interactive visualization of acceleration, velocity, and position via time-series graphs, as well as trajectory plots as needed. These visual representations help students recognize the relationships among physical quantities and identify characteristic features of motion. All processing is performed within the browser, maintaining consistency with the web-based measurement application.

By streamlining numerical integration and visualization, the analysis application reduces cognitive and technical barriers, facilitating a more efficient use of limited class time. As a result, students are able to devote greater attention to discussing experimental results, comparing them with theoretical expectations, and reflecting on the physical principles involved. This web-based approach supports the educational goal of emphasizing conceptual understanding over procedural complexity in mechanics experiments. In the following section, we present measurement examples and analysis results, all of which were generated using this web-based analysis application.

\section{Measurement examples}\label{measurement_examples}
In this section, we evaluate the accuracy of our application using several common types of motion as examples. Note that, during measurement, we activate the sensors and begin data recording while the smartphone is initially at rest. This procedure allows the initial velocity to be set to zero when performing numerical integration to obtain velocity at subsequent time step, which simplifies the analysis and improves reliability. All measurement data presented in this section were obtained using an iPhone 13.

\subsection{Sliding motion on a horizontal rough surface}
A smartphone is placed at rest on a horizontal surface, such as a laboratory bench or desk. After activating the sensors in the measurement application, an initial velocity is imparted by hand, and the smartphone is allowed to slide on the surface. Its traveling direction is defined as the $x$-axis. If the camera module of the smartphone protrudes from the back surface, the device can be attached to a piece of cardboard cut to an appropriate size to ensure smooth sliding.

\begin{figure}
  \begin{center}
    \includegraphics[width=0.8\linewidth]{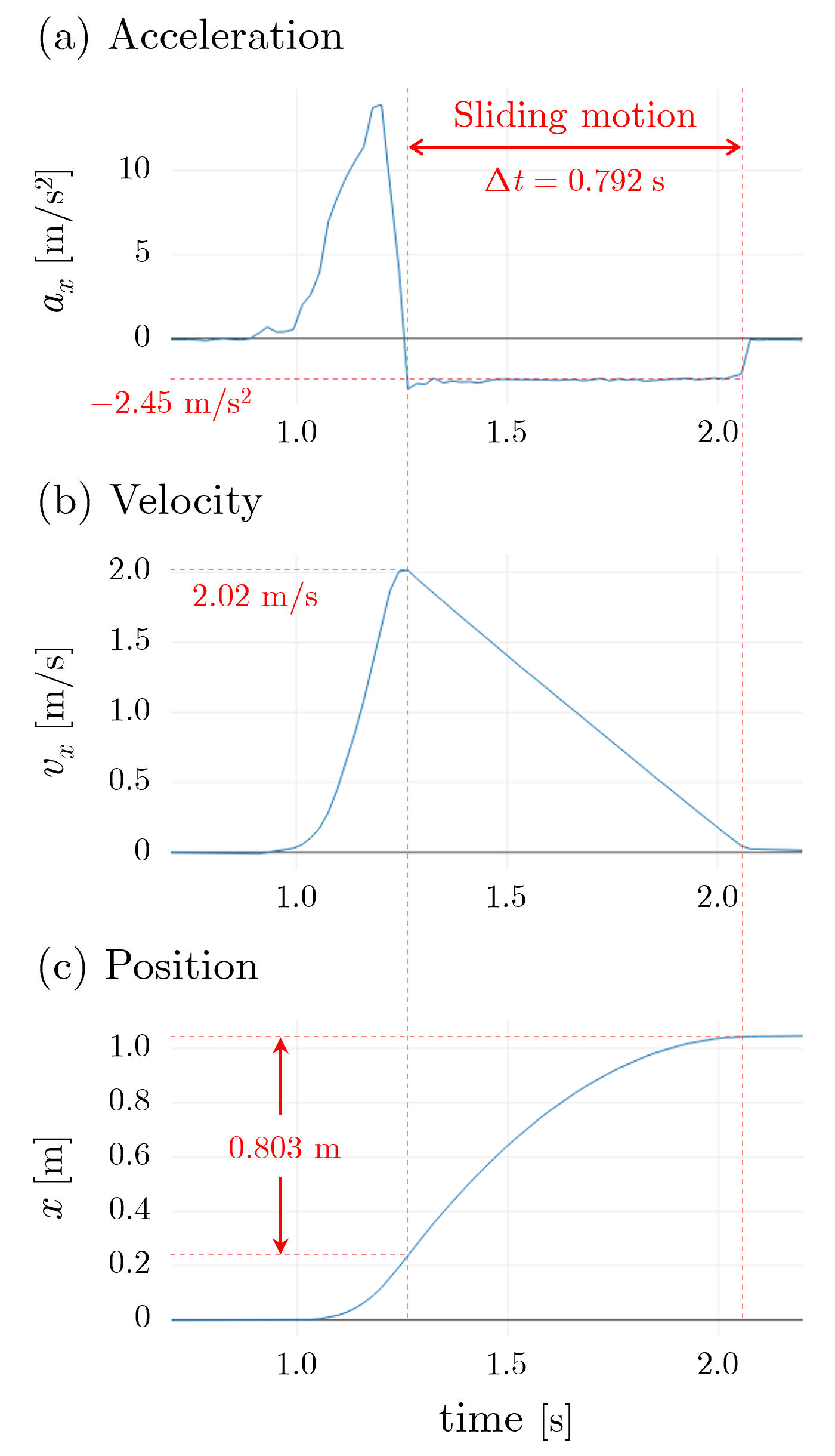}
    \caption{Acceleration, velocity, and position for sliding motion on a horizontal surface under kinetic friction: (a) Measured acceleration in the stationary coordinate system. (b) Velocity obtained by numerical integration. (c) Position obtained by further integration.}
    \label{fig:sliding_motion_on_horizontal_rough_surface}
  \end{center}
\end{figure}

Figure \ref{fig:sliding_motion_on_horizontal_rough_surface}(a) shows the measured acceleration in the stationary coordinate system, while Figs. \ref{fig:sliding_motion_on_horizontal_rough_surface}(b) and \ref{fig:sliding_motion_on_horizontal_rough_surface}(c) show the velocity and position, respectively, obtained through numerical integration. As seen in the acceleration-time plot, the positive acceleration observed between $t = 1.0$ s and 1.25 s corresponds to the impulse generated when the smartphone is pushed by hand. The subsequent, nearly constant negative acceleration of $-2.45$ m/s$^2$ is attributed to kinetic friction acting on the device. From the velocity-time plot in Fig. \ref{fig:sliding_motion_on_horizontal_rough_surface}(b), the velocity at the moment of release is determined to be 2.02 m/s, decreasing linearly with time while the smartphone is sliding. The position-time plot in Fig. \ref{fig:sliding_motion_on_horizontal_rough_surface}(c) indicates a total displacement of 0.803 m from release to rest.

These results clearly demonstrate that the motion corresponds to uniformly accelerated linear motion. Even if the smartphone undergoes slight rotation while sliding, our analysis remains valid because the acceleration is expressed in the stationary global coordinate system. This highlights the effectiveness of the rotation compensation implemented in our application.

Assuming the magnitude of gravitational acceleration to be $g = 9.8$ m/s$^2$, the kinetic friction coefficient $\mu_k$ can be estimated from the measured constant acceleration as
\begin{equation}
\mu_k = \frac{|a|}{g} = \frac{2.45}{9.8} \approx 0.25.
\end{equation}
This value is reasonable for a smartphone sliding on a desk surface, indicating that the measured acceleration is physically consistent.

Based on the recorded acceleration profile, the sliding duration is determined to be $\Delta t = 0.792$ s. Substituting the measured initial velocity and the constant acceleration into the equation for uniformly accelerated linear motion, the theoretical displacement is calculated as
\begin{eqnarray}
x &=& \frac{1}{2} a (\Delta t)^2 + v_0 \Delta t = \frac{1}{2}(-2.45)(\Delta t)^2 + 2.02\Delta t\nonumber\\
&\approx& 0.826\ \mathrm{m}.
\end{eqnarray}
Comparing this value with the displacement obtained from numerical integration (0.803 m), the relative difference is approximately 2.8 $\%$. This level of agreement demonstrates that the proposed measurement/analysis system achieves sufficient accuracy for introductory mechanics experiments. This example enables students to connect measured acceleration data with fundamental kinematic equations and to compare experimental results with theoretical predictions.

\subsection{Projectile motion}
\begin{figure}
  \begin{center}
    \includegraphics[width=0.7\linewidth]{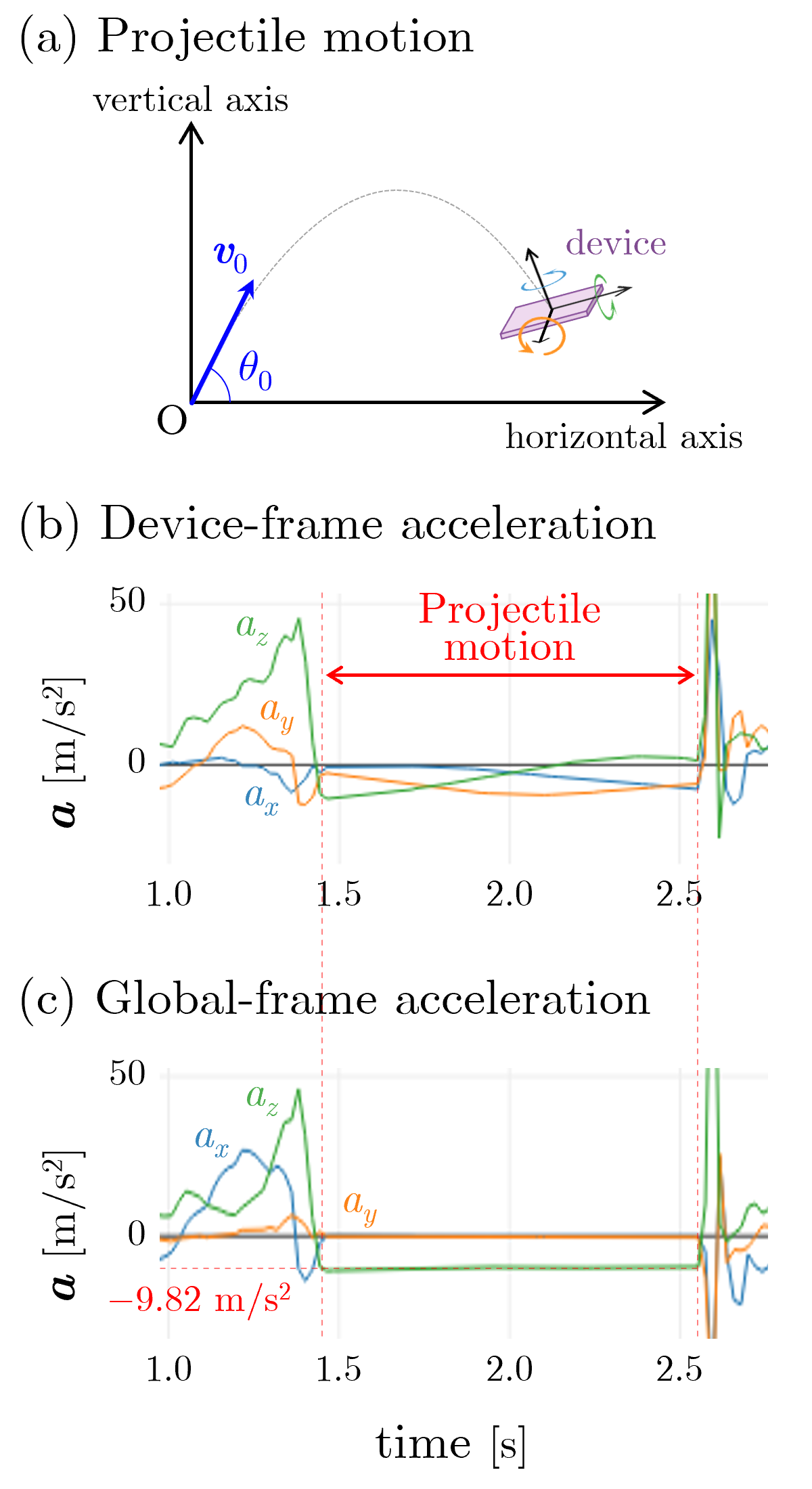}
    \caption{(a) Schematic diagram of projectile motion. (b) Device-frame acceleration. (c) Global-frame acceleration.}
    \label{fig:projectile_motion1}
  \end{center}
\end{figure}

As shown in Fig. \ref{fig:projectile_motion1}(a), the smartphone is thrown by hand to undergo projectile motion. Figures \ref{fig:projectile_motion1}(b) and (c) show the acceleration measured in the device frame and the global frame, respectively. In the global frame, the $z$-axis corresponds to the vertical direction, and the $xy$-plane represents the horizontal plane.

The time interval during which the smartphone undergoes projectile motion spans from $t = 1.44$ s to 2.55 s. The large acceleration observed before $t = 1.44$ s is attributed to the force applied by hand during the launch, while the abrupt change in acceleration after $t = 2.55$ s corresponds to the impact when the smartphone is caught.

Neglecting air resistance, a body in projectile motion is subject only to a constant gravitational acceleration acting vertically downward. As shown in the device-frame acceleration plot in Fig. \ref{fig:projectile_motion1}(b), the smartphone rotates during the projectile motion, causing the vertical direction to shift among the device-frame $x$-, $y$-, and $z$-axes. In contrast, in the global-frame acceleration shown in Fig. \ref{fig:projectile_motion1}(c), the $z$-axis always represents the vertical direction. As a result, only the $z$-component of acceleration has a non-zero value, with $a_z \approx -9.82$ m/s$^2$, which is nearly equal to the magnitude of the gravitational acceleration $g$. This effectively demonstrates that the acceleration in the device-fixed coordinate system is correctly transformed into that in the stationary coordinate system through the rotation compensation.

\begin{figure}
  \begin{center}
    \includegraphics[width=0.8\linewidth]{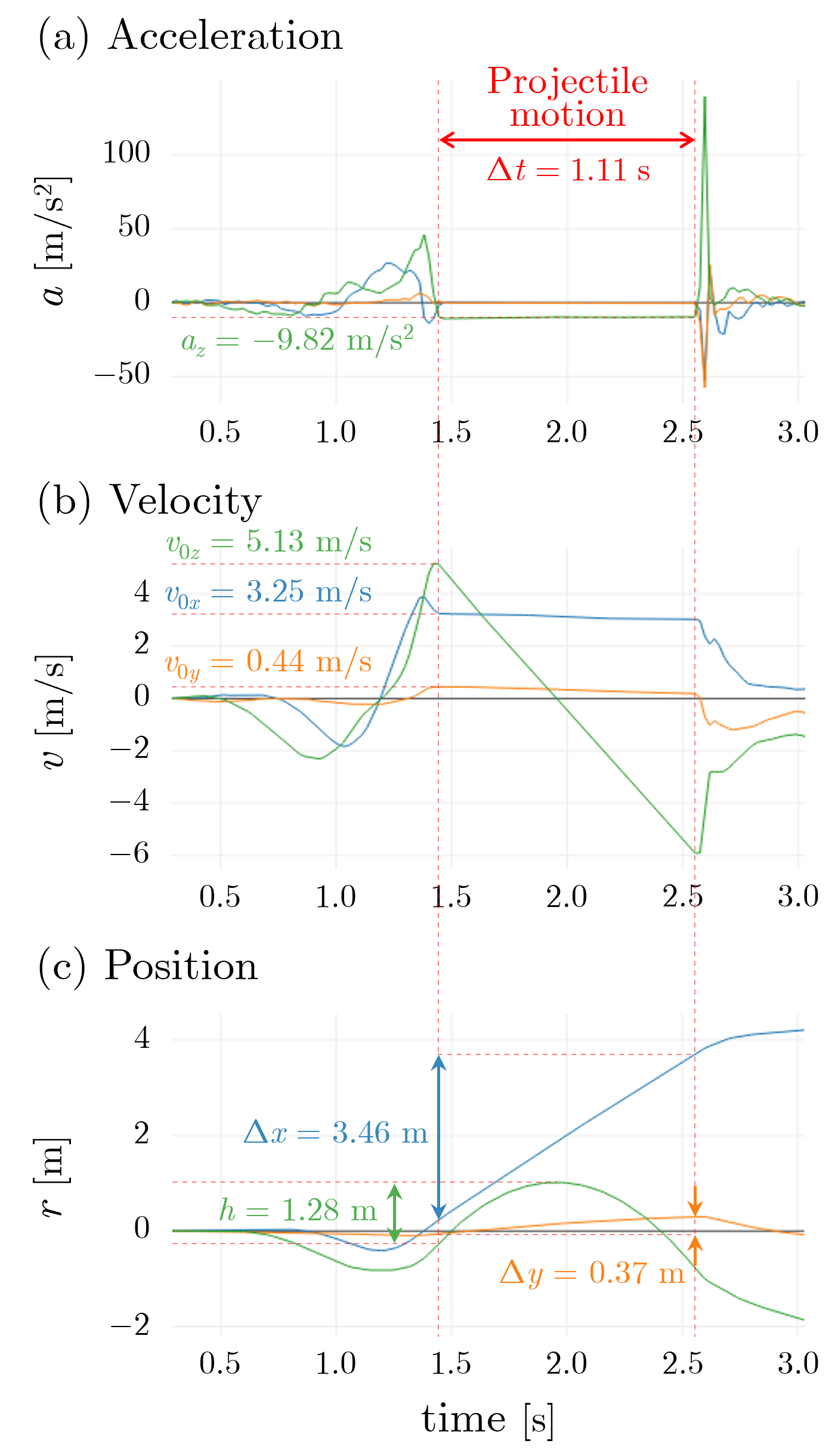}
    \caption{Acceleration, velocity, and position for projectile motion: (a) Measured acceleration in the stationary coordinate system. (b) Velocity obtained by numerical integration. (c) Position obtained by further integration.}
    \label{fig:projectile_motion2}
  \end{center}
\end{figure}

Figure \ref{fig:projectile_motion2}(a) shows the acceleration in the stationary coordinate system, while Figs. \ref{fig:projectile_motion2}(b) and \ref{fig:projectile_motion2}(c) show the velocity and position, respectively, obtained through numerical integration. From the velocity-time plot, the initial velocity of the projectile motion at the moment the smartphone is released is determined to be $\Vec{v}_{0} = (3.25,\ 0.44,\ 5.13)$ m/s. The velocity components in the $x$- and $y$-directions (horizontal) remain approximately constant, while the velocity in the $z$-direction decreases linearly with time, indicating uniformly accelerated motion under gravity.

From the initial velocity components, the launch angle $\theta_0$ is obtained as
\begin{equation}
\theta_0 = \arctan\!\left(\frac{v_{0z}}{\sqrt{v_{0x}^2 + v_{0y}^2}}\right) \approx 57.4^\circ .
\end{equation}
Additionally, the position-time plot in Fig. \ref{fig:projectile_motion2}(c) allows for the determination of the maximum height $h$ relative to the launch point, as well as the displacements $\Delta x$ and $\Delta y$ until the smartphone is caught. The horizontal range $L$ is then calculated as
\begin{equation}
L = \sqrt{(\Delta x)^2 + (\Delta y)^2} \approx 3.48\ \mathrm{m}.
\end{equation}

\begin{figure}
  \begin{center}
    \includegraphics[width=0.75\linewidth]{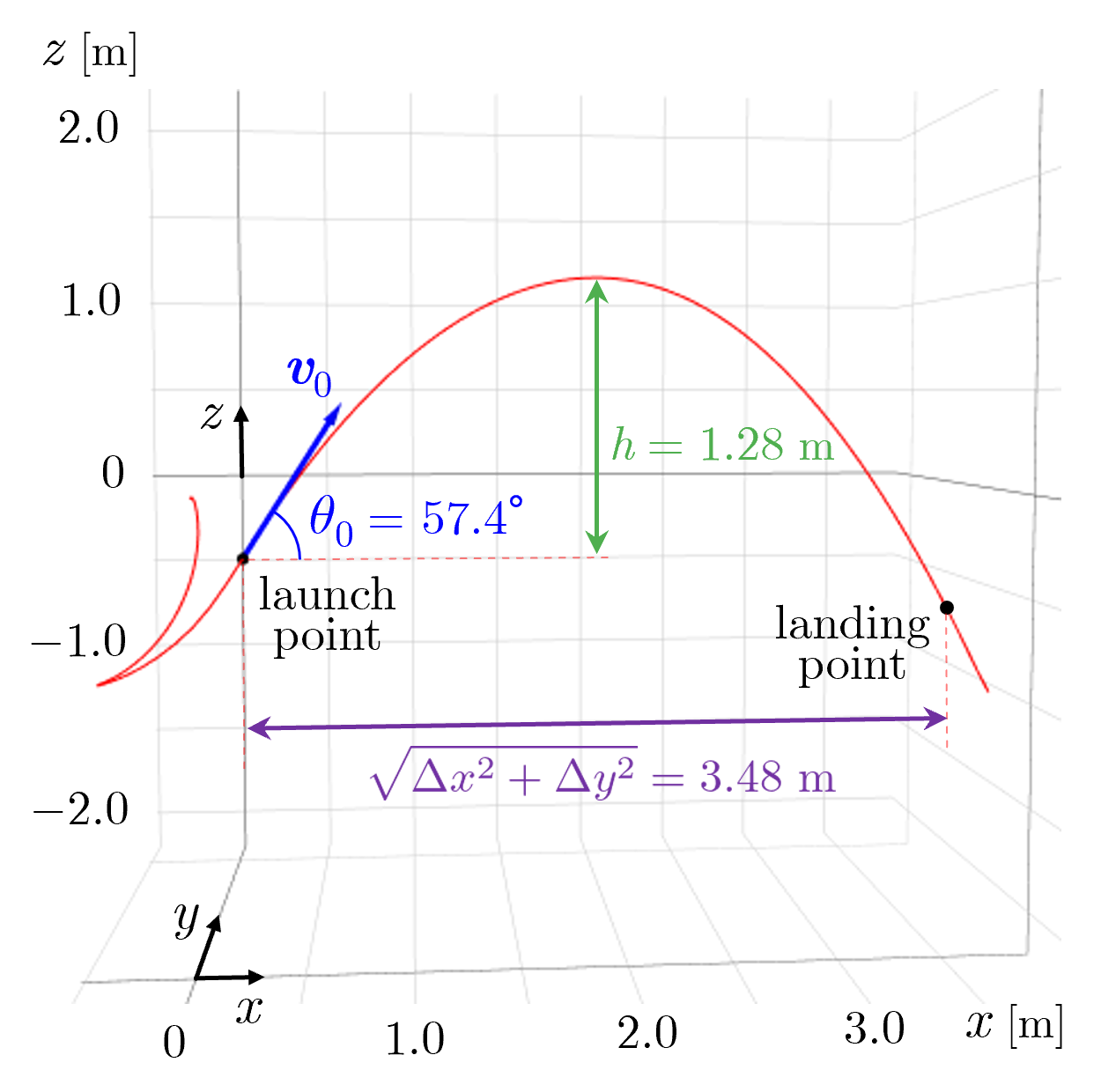}
    \caption{Trajectory of projectile motion.}
    \label{fig:projectile_motion3}
  \end{center}
\end{figure}

Figure \ref{fig:projectile_motion3} shows the trajectory of the motion obtained from the position data. The figure clearly illustrates the smooth parabolic trajectory characteristic of projectile motion, following the initial launch phase where the smartphone is swung by hand.

The experimental results can also be compared with the theoretical predictions for ideal projectile motion. Neglecting air resistance, the maximum height relative to the launch point is expressed as $h_{\mathrm{max}} = v_{0z}^2/2g$. Substituting $v_{0z} = 5.13$ m/s and $g = 9.82$ m/s$^2$, we obtain $h_{\mathrm{max}}\approx1.34$ m. This is consistent with the observed value of 1.28 m in Fig. \ref{fig:projectile_motion2}(c), supporting the validity of the measurement and analysis.

Notably, if the acceleration data were analyzed directly in the device-fixed coordinate system without rotation compensation, the gravitational acceleration would be distributed across different axes as the smartphone rotates. In such a case, numerical integration would yield physically incorrect velocity and position data, obscuring the characteristic features of projectile motion. This example demonstrates that rotation compensation is essential for correctly analyzing three-dimensional motion involving device rotation. Comparing device-frame and global-frame data also highlights the role of coordinate systems in describing vector quantities.

\subsection{Uniform circular motion}
\begin{figure}
  \begin{center}
    \includegraphics[width=0.7\linewidth]{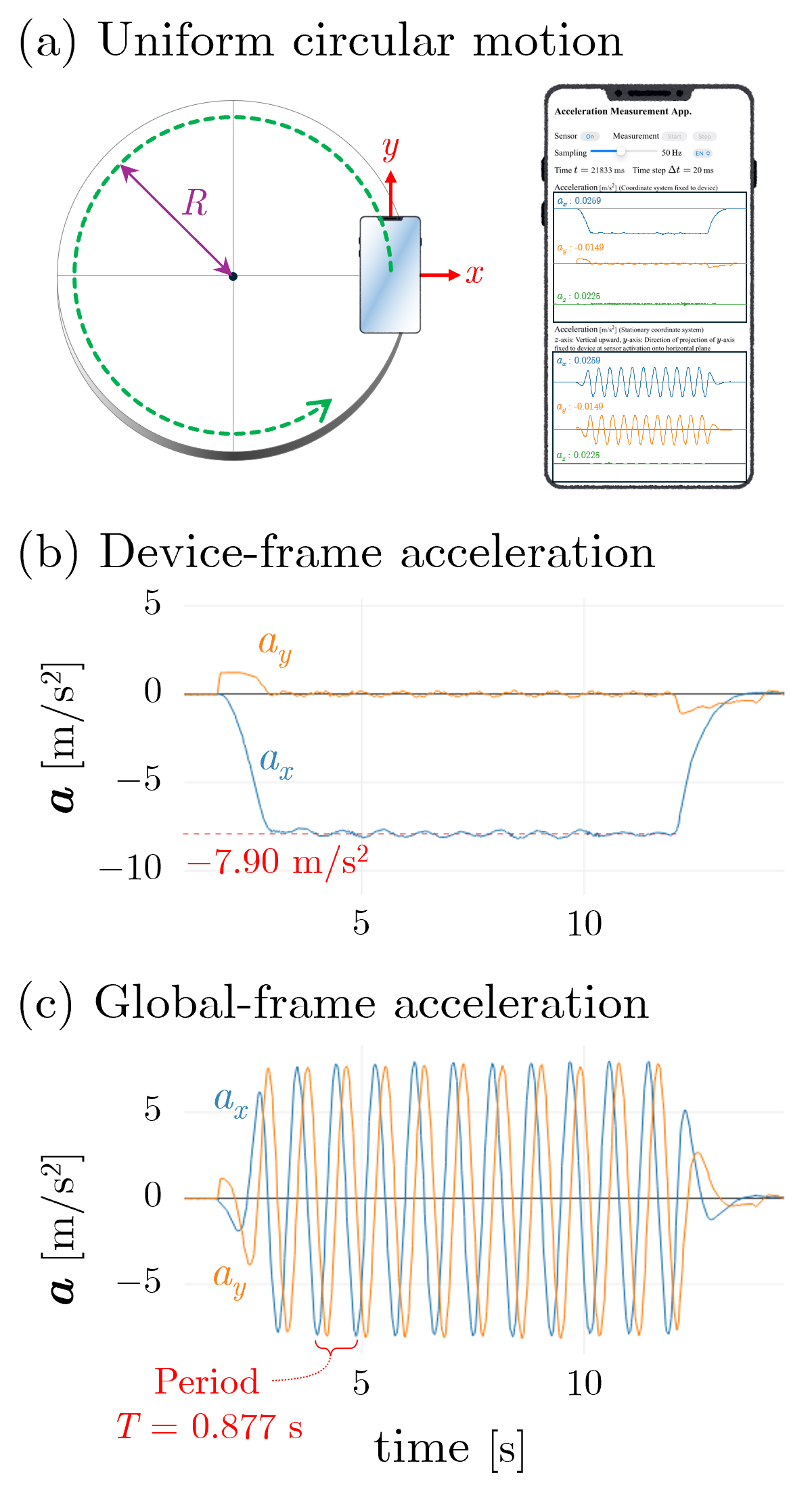}
    \caption{(a) Schematic diagram of circular motion and measurement screen. (b) Device-frame acceleration. (c) Global-frame acceleration.}
    \label{fig:circular_motion1}
  \end{center}
\end{figure}

As shown in Fig. \ref{fig:circular_motion1}(a), a smartphone is fixed to a rotating platform placed on a horizontal surface. The platform is driven by an electric motor to rotate at a constant angular velocity, causing the smartphone to undergo uniform circular motion with a radius $R$. Since the exact location of the sensors within the smartphone is not publicly specified, we define the radius $R$ as the distance from the center of the rotating platform to the geometric center of the smartphone, which is approximately 16 cm in our setup.

Figures \ref{fig:circular_motion1}(b) and (c) show the acceleration measured in the device frame and the global frame, respectively. Since the smartphone performs circular motion in the horizontal plane ($xy$-plane), only the $x$- and $y$-components of acceleration are shown. In the device-frame acceleration [Fig. \ref{fig:circular_motion1}(b)], although some vibrational components are visible, a centripetal acceleration of nearly constant magnitude appears primarily along the $x$-axis, with a value of approximately 7.90 m/s$^2$. In contrast, the $x$- and $y$-components of the global-frame acceleration [Fig. \ref{fig:circular_motion1}(c)] exhibit clean sinusoidal waveforms with a phase difference of $\pi/2$, which is characteristic of uniform circular motion.

In Fig. \ref{fig:circular_motion1}(c), the stable rotation of the smartphone is observed approximately from $t = 4$ s to 11 s, where the period of the circular motion is determined to be about 0.877 s. The amplitude of the acceleration is approximately equal to the magnitude of the centripetal acceleration, 7.90 m/s$^2$.

\begin{figure}
  \begin{center}
    \includegraphics[width=0.7\linewidth]{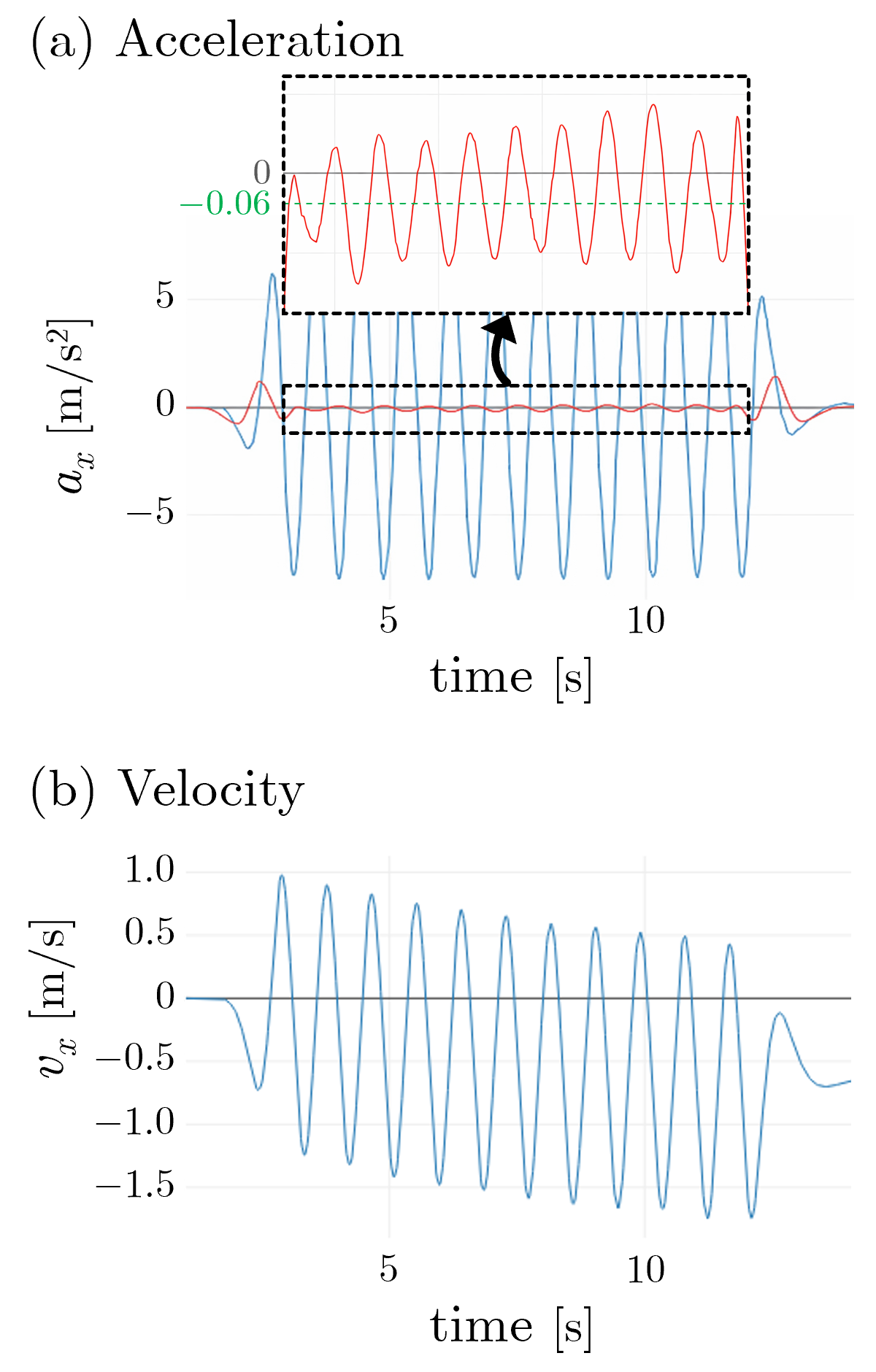}
    \caption{(a) $x$-component of acceleration (blue line) and its moving average with a time interval of 0.877 s (red line). The inset figure shows a vertically enlarged graph of the moving average. (b) $x$-component of velocity obtained by numerically integrating the raw data of the $x$-component of acceleration.}
    \label{fig:circular_motion2}
  \end{center}
\end{figure}

Figure \ref{fig:circular_motion2}(a) shows the $x$-component of the global-frame acceleration together with its moving average over a time window corresponding to one period (red line), and Fig. \ref{fig:circular_motion2}(b) shows the result of simple numerical integration of this acceleration data. Although the acceleration data in Fig. \ref{fig:circular_motion2}(a) appears to oscillate sinusoidally around zero, the moving average reveals a slight offset of the oscillation center toward the negative side (approximately -0.06 m/s$^2$). Due to this small bias in the acceleration data, which is presumably attributable to sensor measurement errors, the velocity obtained through numerical integration accumulates the offset over time, leading to a gradual drift in the negative direction, as shown in Fig. \ref{fig:circular_motion2}(b). Such behavior, known as integration drift, is physically incorrect for an object in uniform circular motion.

\begin{figure}
  \begin{center}
    \includegraphics[width=0.8\linewidth]{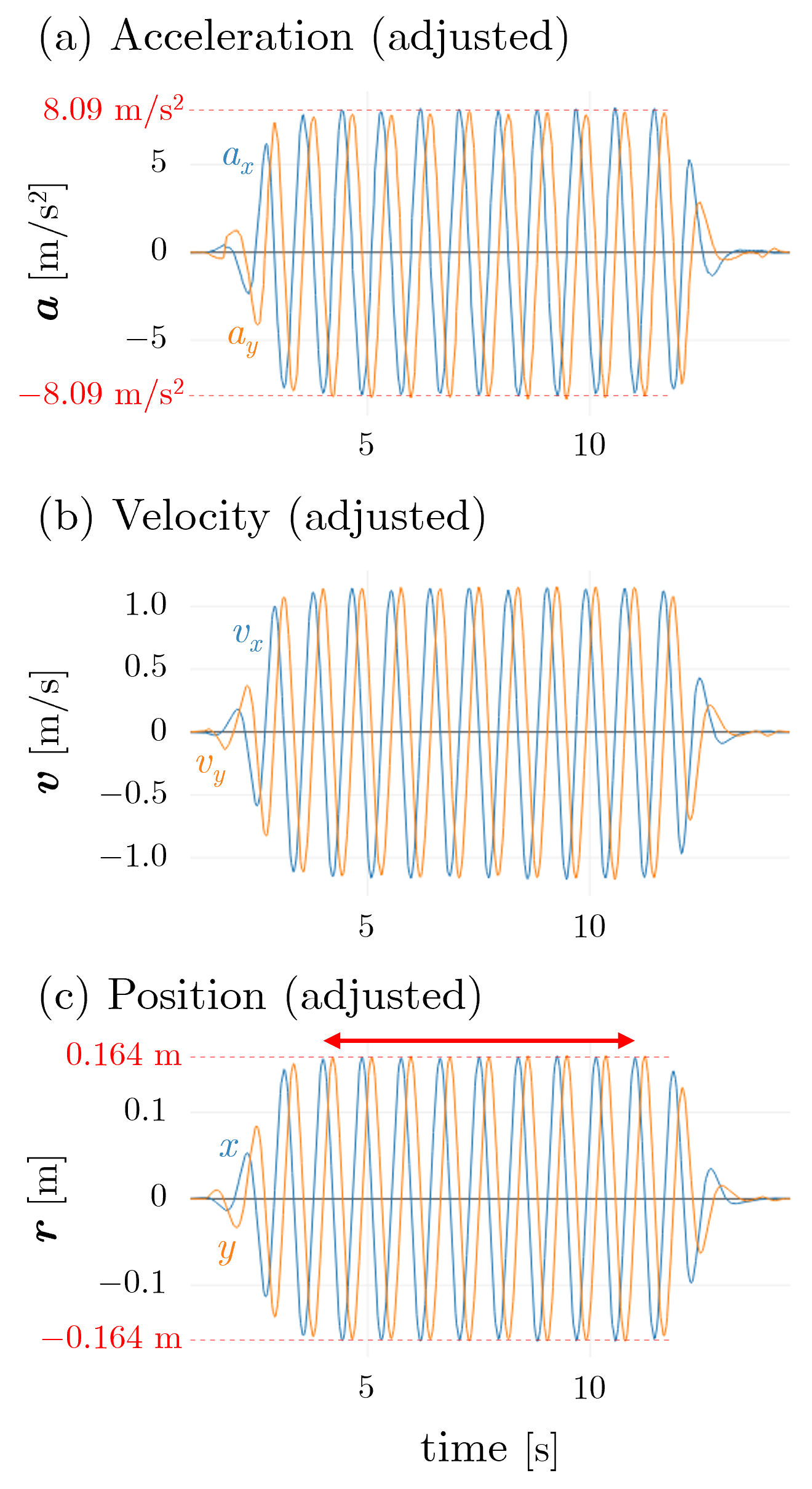}
    \caption{Acceleration, velocity and position for uniform circular motion, which are adjusted by applying the offset correction: (a) Adjusted acceleration. (b) Adjusted velocity. (c) Adjusted position.}
    \label{fig:circular_motion3}
  \end{center}
\end{figure}

In the analysis of oscillatory motions, it is therefore necessary to adjust the data so that the center of oscillation is zero. An effective approach is to subtract the moving-average component from the measured data, which can be performed directly using our analysis application. Figure \ref{fig:circular_motion3}(a) shows the results after applying this offset correction to the acceleration data, while Figs. \ref{fig:circular_motion3}(b) and \ref{fig:circular_motion3}(c) present the velocity and position, respectively, obtained through numerical integration and the offset correction. The moving-average window is selected to match the measured rotational period (0.877 s) so that the center of oscillation is zero without significantly affecting the oscillation signal. During the interval of stable uniform circular motion ($t = 4$ s to 11 s, indicated by the double-headed arrow in Fig. \ref{fig:circular_motion3}(c)), the acceleration, velocity, and position agree well with the theoretical behavior of uniform circular motion. Notably, in the adjusted acceleration, velocity, and position plots shown in Fig. \ref{fig:circular_motion3}, the behavior during the startup phase ($t < 4$ s) and the deceleration phase ($t > 11$ s) deviates slightly from the actual physical motion. As seen in Fig. \ref{fig:circular_motion3}(c), the time dependence of $r = \sqrt{x^2 + y^2}$ indicates that the radius of the device's circular motion is initially zero and gradually increases until $t \approx 4$ s, at which point stable rotation is established. Although this behavior is an artifact resulting from the data adjustment method employed in our application, the analysis of the uniform circular motion itself remains valid, as it relies exclusively on the data obtained during the stable rotation interval ($4\ {\rm s} < t < 11\ {\rm s}$).

\begin{figure}
  \begin{center}
    \includegraphics[width=0.75\linewidth]{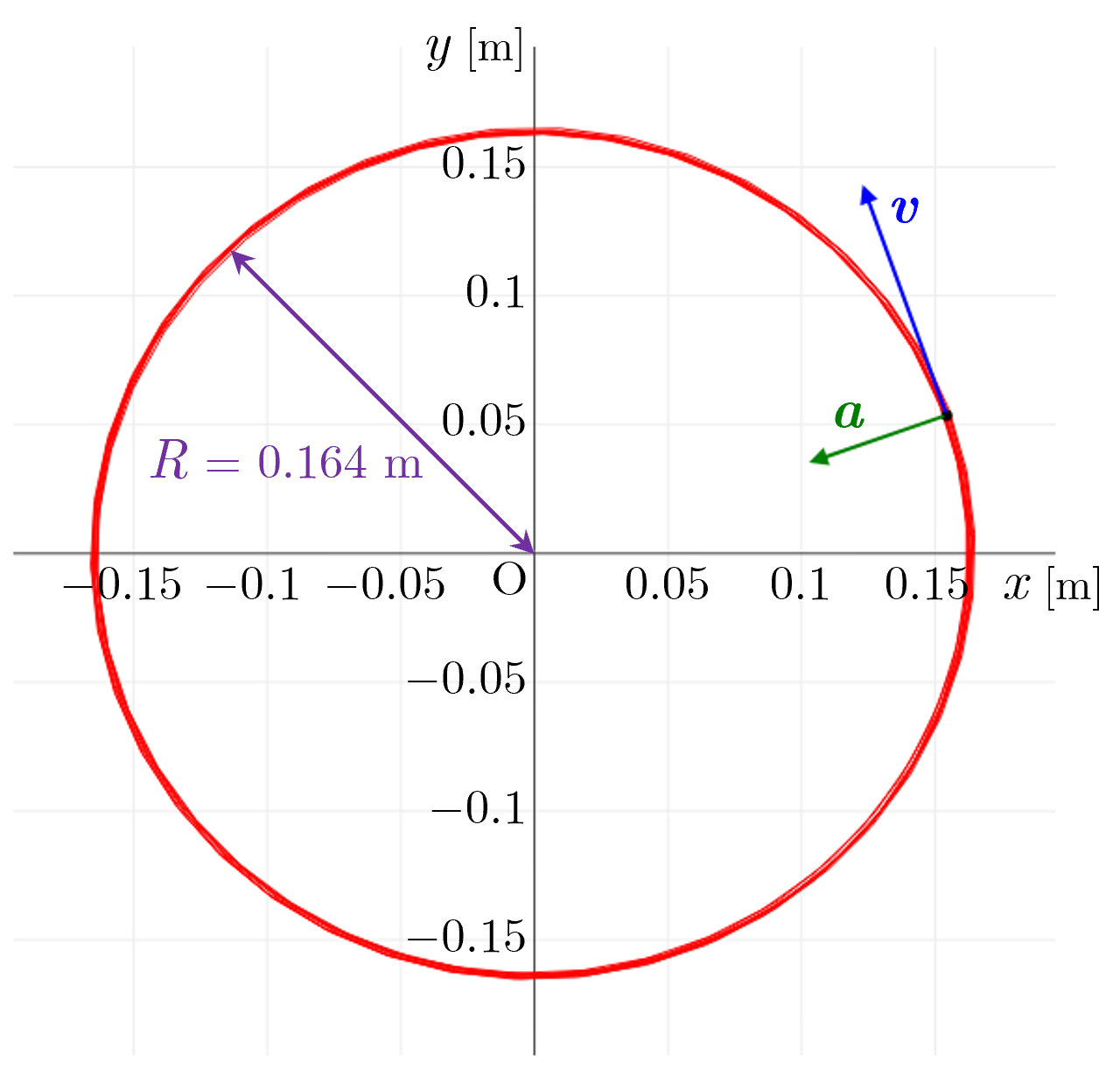}
    \caption{Trajectory of circular motion.}
    \label{fig:circular_motion4}
  \end{center}
\end{figure}

Figure \ref{fig:circular_motion4} shows the trajectory of the circular motion over eight periods within the stable rotation interval indicated in Fig. \ref{fig:circular_motion3}(c). Although a small spread in the circular path is observed, the trajectory forms a nearly perfect circle with a radius of approximately 16 cm, demonstrating that the motion can be analyzed with reasonably high accuracy.

From Fig. \ref{fig:circular_motion3}(a), the peak amplitude of the acceleration is approximately 8.09 m/s$^2$, as indicated by the dashed red lines. Based on the measured period $T \simeq 0.877$ s, the angular velocity $\omega$ is estimated as
\begin{equation}
\omega = \frac{2\pi}{T} \simeq 7.16\ \mathrm{rad/s}.
\end{equation}
Using the theoretical relation for uniform circular motion, $a = R\omega^2$, and the measured radius $R = 0.164$ m, the magnitude of the theoretical centripetal acceleration is
\begin{equation}
a_{\mathrm{th}} = R\omega^2 \simeq 8.4\ \mathrm{m/s^2},
\end{equation}
which is in reasonable agreement with the measured value of 8.09 m/s$^2$ within experimental uncertainty.

The slight offset revealed by the moving average becomes significant after numerical integration because integration accumulates low-frequency components over time. This integration drift is clearly seen in Fig. \ref{fig:circular_motion2}(b). Subtracting the moving-average component effectively suppresses the drift and restores velocity and position data consistent with theoretical expectations.

The observed offset in the acceleration data can be attributed to several possible sources of experimental error. One major factor is the limited accuracy and stability of the smartphone's built-in accelerometer, which may exhibit bias drift or sensitivity variations over time. In addition, slight misalignment between the assumed global coordinate system and the actual orientation of the device can introduce residual constant components in the measured acceleration. Furthermore, imperfections in the experimental setup, such as a small tilt of the rotating platform, mechanical vibrations from the motor, or slight deviations of the smartphone's center of mass from the rotation axis, can also contribute to the observed offset. These factors are difficult to eliminate completely in a simple educational experiment, but their influence can be effectively mitigated through appropriate data processing.

\section{Classroom implementation}
As demonstrated by the measurement examples in the previous section, the proposed applications enable the measurement and analysis of acceleration in the global frame with sufficient accuracy for various typical motions. Based on these results, a group-based learning activity using these applications was conducted in a second-year undergraduate mechanics course at our university. This activity followed lecture-based instruction covering one- and two-dimensional motion, projectile motion, circular motion, and Newton's laws of motion.

During the activity, students measured not only the standard types of motion described previously but also phenomena of their own choosing. They were instructed to begin measurements from rest, avoid rapid rotation of the smartphone, and refrain from motions involving abrupt changes in acceleration. In similar group activities conducted prior to the introduction of these applications, students often paid little attention to device rotation, resulting in analytical results that failed to correspond to the actual motion. In contrast, when using the current applications, no cases were observed where the results were entirely inconsistent with the measured motion. The ability to immediately visualize analysis results was particularly beneficial in supporting students' interpretation and discussion during the activity.

\begin{figure*}
  \begin{center}
    \includegraphics[width=0.7\linewidth]{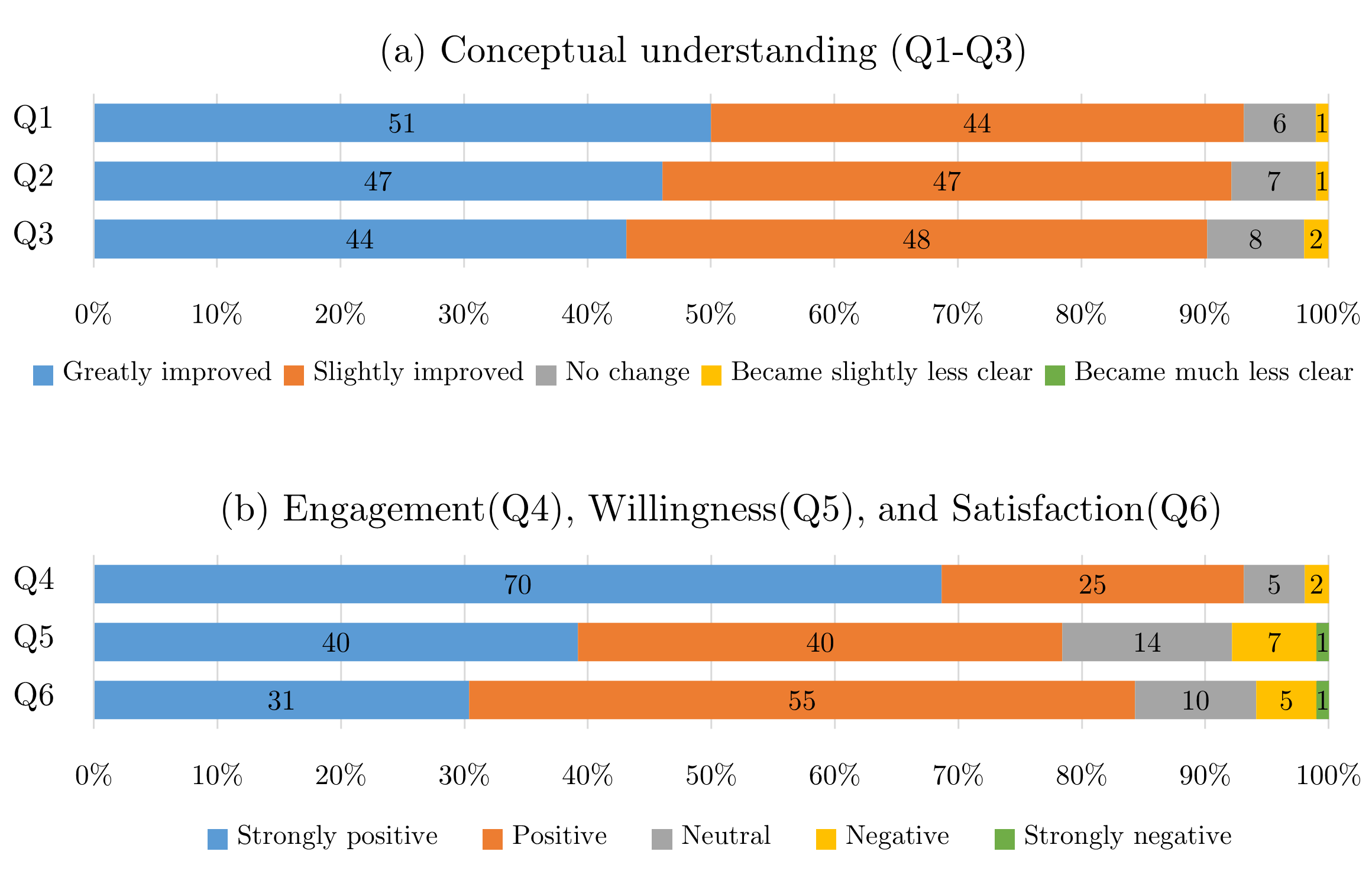}
    \caption{(a) Distribution of students' responses to questionnaire items on conceptual understanding (Q1–Q3). The horizontal stacked bar charts show the percentage of responses for each Likert-scale category. (b) Distribution of students' responses regarding engagement, willingness, and overall satisfaction (Q4–Q6).}
    \label{fig:questionnaire_result}
  \end{center}
\end{figure*}

A questionnaire survey was conducted following the activity, with responses obtained from 102 students ($n = 102$). Questions Q1–Q6 utilized a Likert scale to evaluate students' conceptual understanding, engagement, and perceptions of the smartphone-based experiment. The response distributions are summarized in Fig. \ref{fig:questionnaire_result}.

Overall, the questionnaire results indicate strong educational effectiveness of the activity. More than 90\% of students reported an improved understanding of the relationships among position, velocity, and acceleration (Q1), as well as of the process of deriving velocity and position through numerical integration (Q2). Similarly, approximately 90\% of the students noted a clearer connection between the experimental activity and the lecture content (Q3), suggesting that the activity effectively bridged theoretical concepts and real-world observations.

Student engagement was also high: over 90\% of respondents indicated active or moderately active participation in the group work (Q4). Furthermore, a large majority expressed positive attitudes toward future smartphone-based activities (Q5) and reported high overall satisfaction with the learning experience (Q6). These results demonstrate that combining smartphone-based measurements with immediate web-based analysis can promote conceptual understanding, active participation, and positive learning experiences in introductory mechanics courses.

In addition to the Likert-scale items, students were asked to describe in their own words what they had learned through the group-learning activity (Q7). These open-ended responses were analyzed qualitatively and classified into six thematic categories. Table \ref{tab:categories_q7} summarizes these categories, their descriptions, and the number of comments associated with each category (multiple codes per response were permitted).

\begin{table*}[tb]
  \caption{Categorization of students' open-ended responses to Q7 (multiple coding allowed, $n = 102$).}
  \label{tab:categories_q7}
  \centering
  \small
  \begin{tabular}{l l r}
    \hline
    Category & Description & Frequency \\
    \hline

    \parbox[t]{3.8cm}{Data analysis and visualization skills} &
    \parbox[t]{6.8cm}{Graph creation, numerical integration, and use of analysis tools (e.g., Excel, data analysis apps)} &
    57 \\[-0.3em]
    \multicolumn{3}{c}{\dotfill} \\

    \parbox[t]{3.8cm}{Collaborative and communication skills} &
    \parbox[t]{6.8cm}{Group discussion, role sharing, teamwork, and learning through peer interaction} &
    36 \\[-0.3em]
    \multicolumn{3}{c}{\dotfill} \\

    \parbox[t]{3.8cm}{Educational impact of familiar devices} &
    \parbox[t]{6.8cm}{Increased engagement and accessibility through the use of smartphones as familiar measurement tools} &
    35 \\[-0.3em]
    \multicolumn{3}{c}{\dotfill} \\

    \parbox[t]{3.8cm}{Inquiry-based learning attitudes} &
    \parbox[t]{6.8cm}{Developing the ability to make predictions, ask ``why'' questions, and reflect on unexpected results} &
    26 \\[-0.3em]
    \multicolumn{3}{c}{\dotfill} \\

    \parbox[t]{3.8cm}{Connection between theory and measurement} &
    \parbox[t]{6.8cm}{Understanding the relationship between formulas and real-world motion based on measured data} &
    23 \\[-0.3em]
    \multicolumn{3}{c}{\dotfill} \\

    \parbox[t]{3.8cm}{Awareness of error and idealization} &
    \parbox[t]{6.8cm}{Recognition of discrepancies between theoretical predictions and measured data, including measurement errors and modeling assumptions} &
    9 \\
    \hline
  \end{tabular}
\end{table*}

The most frequently mentioned category was \emph{data analysis and visualization skills} (57 comments), including graph creation, numerical integration, and the use of spreadsheet software or dedicated analysis applications. Many students reported that visualizing acceleration, velocity, and position data helped them understand the relationships among these physical quantities.

A substantial number of responses also emphasized \emph{collaborative and communication skills} (36 comments), indicating that group discussions, role sharing, and peer interaction played an important role in the learning process. Similarly, the \emph{educational impact of familiar devices}, such as smartphones, was frequently noted (35 comments), suggesting that accessibility and familiarity contributed to increased motivation and engagement.

Comments regarding the \emph{connection between theoretical formulas and measured motion} (23 comments) and \emph{inquiry-based learning attitudes} (26 comments) indicate that the activity supported conceptual understanding beyond procedural problem solving. Students described experiences of making predictions, questioning unexpected results, and reflecting on discrepancies between theory and observation. Although less frequent, several students explicitly mentioned \emph{awareness of measurement error and idealization} (9 comments), suggesting that encountering imperfect data encouraged critical thinking about experimental limitations and modeling assumptions.

Overall, the activity effectively connected abstract kinematic concepts with measured motion. Students engaged with physical quantities as experimentally accessible data rather than purely symbolic expressions. The results also highlight the importance of collaborative work and familiar devices in promoting engagement. Future implementations should incorporate clearer instructional scaffolding.

\section{Conclusion}
We presented a web-based acceleration-measurement application featuring a rotation-compensation capability that transforms acceleration data from the device-fixed coordinate system into a stationary global frame. Together with a companion web-based analysis application, this system allows students to measure, visualize, and analyze acceleration, velocity, and position directly using their own smartphones, without requiring any software installation. The web-based design offers significant advantages for classroom use in terms of accessibility and platform independence.

Measurement examples involving representative types of motion, including sliding motion under kinetic friction, projectile motion and uniform circular motion, demonstrated that rotation-compensated acceleration enables physically meaningful analysis even when the device changes orientation. These examples also provided practical opportunities to discuss numerical integration and data-processing issues, such as drift caused by small offsets in raw acceleration data.

The classroom implementation in an undergraduate mechanics course indicated that the applications supported group-based experimental learning and helped students relate measured data to kinematic concepts. Questionnaire responses and qualitative feedback suggested improved understanding of the relationships among acceleration, velocity, and position, as well as positive engagement with smartphone-based experiments. Although the present evaluation is based primarily on self-reported responses, the results suggest positive educational impact; a more systematic assessment of learning gains will be addressed in future work.

Overall, the proposed system reduces technical barriers and allows students to focus on physical interpretation and discussion of motion. It provides a flexible and practical tool for incorporating smartphone-based experiments into physics and engineering education.

\section*{Acknowledgments}
The authors thank Y. Takai for presenting their results prior to publication and valuable discussions. This work was supported by JSPS KAKENHI Grant Number JP21K02811.

\section*{Data availability}
The acceleration data shown in the measurement examples in Section \ref{measurement_examples} is available at the following URL:
https://github.com/natieck/phys\_apps/tree/main/\\
acceleration\_measurement\_and\_analysis/sample\_data

\section*{Ethical statement}
This study targeted undergraduate engineering students enrolled in a second-year mechanics course. Participation in the study was voluntary, and consent was obtained from all participants in April 2024, permitting the use of their anonymized responses for research and publication purposes. No personally identifiable information was collected, and all data were analyzed only in aggregated form.

\end{document}